\documentclass[11pt]{article}

\usepackage{fullpage}
\usepackage{amsmath, amssymb, amsfonts}
\usepackage{graphicx}
\usepackage{hyperref}
\usepackage{cite}
\usepackage{booktabs}
\usepackage{multirow}
\usepackage{algorithm}
\usepackage{algorithmic}
\usepackage{subcaption}

\title{An Optimal Alignment-Driven Iterative Closed-Loop Convergence Framework for High-Performance Ultra-Large Scale Layout Pattern Clustering}

\author{
    Shuo Liu \\
    Southeast University \\
    \texttt{220246703@seu.edu.cn}
}

\date{\today}

\begin{document}

\maketitle

\begin{abstract}
With the aggressive scaling of Very Large Scale Integration (VLSI) technology, the volume of layout patterns has exploded to the order of billions, creating a critical bottleneck for Design for Manufacturability (DFM) applications such as Optical Proximity Correction (OPC). Pattern clustering is essential to reduce this data complexity, yet existing approaches face three fundamental challenges: the computational prohibitiveness of $O(N^2)$ pairwise comparisons, the intractability of locating optimal pattern centers within continuous marker regions (often compromised by sub-optimal discrete sampling), and the difficulty of balancing execution speed with clustering quality.

To address these challenges, this paper proposes an Optimal Alignment-Driven Iterative Closed-Loop Convergence Framework. First, to resolve the center alignment ambiguity, we introduce a hybrid suite of high-performance alignment algorithms: an FFT-based Phase Correlation method for cosine similarity constraints that decouples spatial position from image content, and a Robust Geometric Min-Max strategy for edge displacement constraints that analytically solves for the global optimum in constant time. Second, we model the clustering task as a Set Cover Problem (SCP) and employ a Surprisal-Based Lazy Greedy heuristic to prioritize rare patterns, integrated within a coarse-to-fine iterative refinement loop that ensures convergence to a high-quality solution. Finally, a multi-stage pruning mechanism is implemented to filter over 99

Experimental results on the 2025 China Postgraduate EDA Elite Challenge benchmark demonstrate that our framework achieves a 93.4

\end{abstract}

\section{Introduction}

\subsection{Background and Motivation}
\begin{figure}[t]
    \centering
    \includegraphics[width=0.85\textwidth]{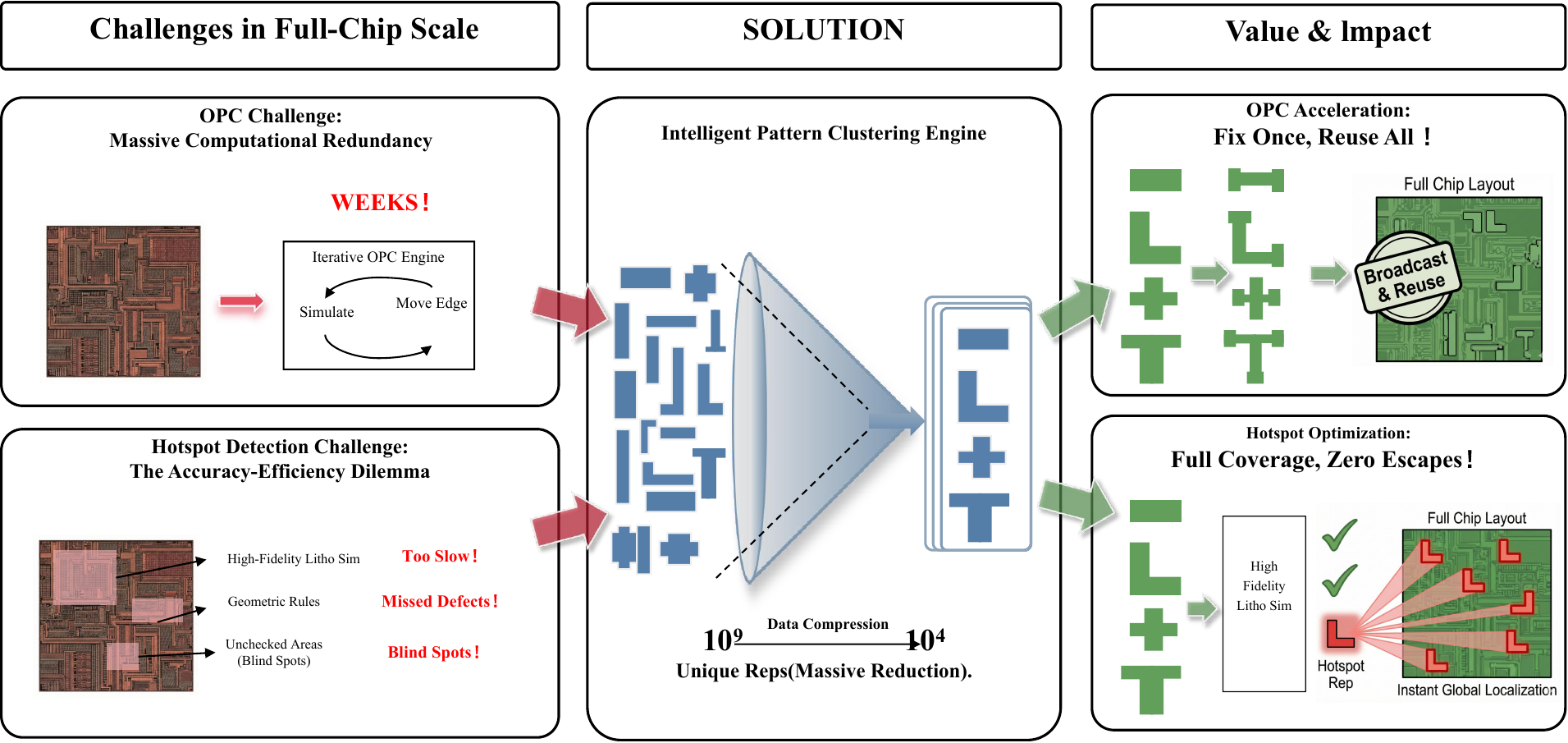}
    \caption{Overview of the critical role of pattern clustering in accelerating VLSI DFM flow.}
    \label{fig:motivation}
\end{figure}
As Very Large Scale Integration (VLSI) technology advances into the 5nm node and beyond, the complexity of layout designs has increased exponentially. The number of layout patterns in a full-chip design can easily reach millions or even billions. This data explosion imposes a tremendous computational burden on Design for Manufacturability (DFM) applications, such as Optical Proximity Correction (OPC), lithography simulation, and hotspot detection \cite{chang2017iclaire, chen2017minimizing}.

To mitigate this challenge, layout pattern clustering has emerged as a fundamental technique. As illustrated in Figure \ref{fig:motivation}, by grouping millions of geometrically identical or highly similar patterns into representative clusters, redundant computations in OPC and hotspot detection can be drastically reduced. This mechanism allows downstream tools to process only a representative pattern from each cluster rather than the entire dataset, which not only accelerates the physical verification cycle but also ensures consistency in corrections across similar local environments \cite{he2023general}.

\subsection{Related Work}
In recent years, numerous algorithms have been proposed to address the pattern clustering problem, which can generally be categorized into density-based, geometry-based, and feature-based methods. Early works, such as \textbf{iClaire} proposed by Chang et al. \cite{chang2017iclaire}, utilized a density-based clustering framework combined with machine learning techniques. While effective for initial screening, iClaire typically relies on a pre-defined grid or limited candidate anchors for alignment. Recognizing that pattern positions are not fixed, Chen et al. \cite{chen2017minimizing} introduced the concept of \textbf{clip shifting}, demonstrating that adjusting the pattern center within a defined marker region can significantly improve matching rates. More recently, He et al. \cite{he2023general} and Lin et al. \cite{lin2023layout} have explored \textbf{geometric matching} strategies. However, these methods predominately employ heuristic search strategies—relying on discrete sampling points or iterative greedy shifts—to approximate the optimal alignment within the continuous solution space.

\subsection{Challenges and Limitations}

\begin{figure*}[t]
    \centering
    \includegraphics[width=0.85\textwidth]{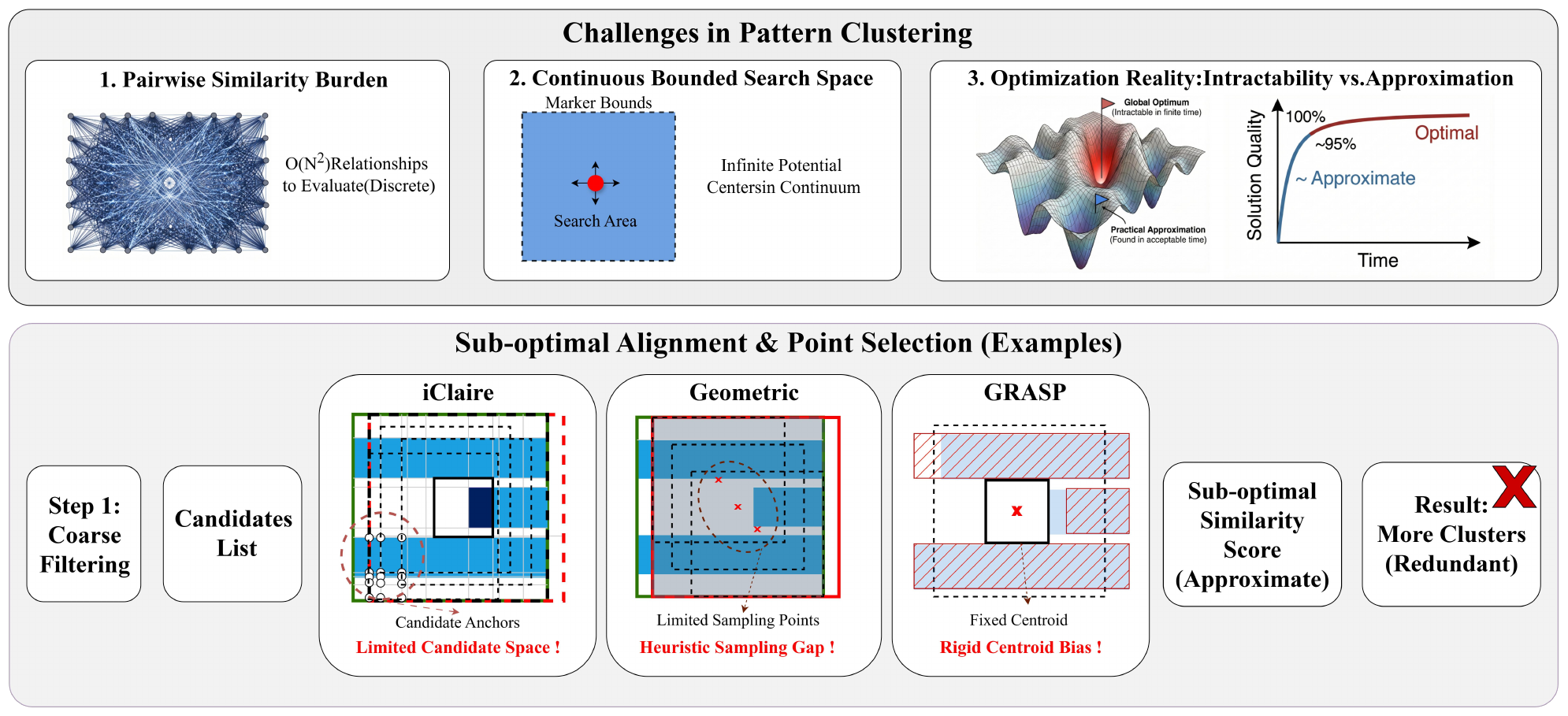} 
    \caption{Challenges and limitations of existing layout clustering methods.}
    \label{fig:challenges}
\end{figure*}

Despite these advancements, existing state-of-the-art solutions still suffer from critical limitations. As visualized in \textbf{Figure \ref{fig:challenges}}, tackling ultra-large-scale layouts under strict alignment constraints presents three fundamental hurdles:

\begin{itemize}
    \item \textbf{Computational Bottleneck:} The pairwise similarity calculation inherently carries an $O(N^2)$ time complexity. With the discrete interaction network growing exponentially, existing algorithms struggle to maintain throughput when $N$ exceeds tens of thousands, often resulting in runtime explosions.

    \item \textbf{Sub-optimal Center Alignment (The Core Pain Point):} A defining characteristic of this problem is the continuous bounded search space. The similarity score is highly sensitive to relative alignment, yet finding the unique global optimum is analytically difficult. 
    
    Currently, heuristics are widely used to approximate this process, but they introduce structural flaws. For instance, \textbf{iClaire} restricts the search to a finite set of discrete anchors, ignoring potential optima between grid points; similarly, \textbf{Geometric} approaches rely on heuristic sampling gaps, leaving blind spots in the search area. These flaws lead to sub-optimal similarity scores (False Negatives) and a consequently inflated number of clusters (Redundancy).

    \item \textbf{Speed-Accuracy Trade-off:} Achieving high precision typically implies intractable computational costs with traditional solvers. Consequently, existing methods are forced to compromise, settling for approximate solutions that sacrifice clustering quality for execution speed—a trade-off our framework aims to eliminate.
\end{itemize}

\subsection{Our Contributions}
To address these challenges, we propose an \textbf{Optimal Alignment–Driven Iterative Closed-Loop Framework}. Unlike prior works that rely on heuristic shifting, our method analytically solves the alignment problem to ensure global optimality. Our major contributions are as follows:

\begin{enumerate}
    \item \textbf{High-Performance Hybrid Alignment Schemes:} 
    We introduce a versatile suite of analytical alignment strategies tailored to different constraint types:
    \begin{itemize}
        \item For \textbf{area-based (cosine) constraints}, we provide two complementary solutions: an \textbf{FFT-based phase correlation} method for maximum accuracy by fully decoupling spatial position from content, and a \textbf{robust geometric min–max strategy} that achieves extreme speed through interval competition with negligible precision loss.
        \item For \textbf{edge displacement constraints}, we propose a specialized \textbf{geometric min–max approximation} that analytically solves for the optimal translation vector using interval balancing.
    \end{itemize}

    \item \textbf{Iterative Closed-Loop Framework:} 
    We design a coarse-to-fine iterative architecture that integrates pre-screening, rough clustering, and rigorous refinement. Patterns that fail in earlier stages are re-inserted with tightened constraints, ensuring convergence toward a minimal set of clusters.

    \item \textbf{Surprisal-Based SCP Solver with Lazy Acceleration:} 
    We formulate the clustering process as a Set Cover Problem (SCP). Inspired by the information-theoretic metric proposed by Adamo et al.~\cite{adamo2023surprisal}, we adopt ``surprisal’’ to prioritize rare or high-information patterns.  
    To scale this approach to ultra-large VLSI datasets, we incorporate a \textbf{lazy greedy evaluation} mechanism, which avoids redundant score updates and dramatically reduces solver complexity. This allows millisecond-level convergence on million-scale datasets.

    \item \textbf{Parallelism and System Optimization:} 
    Through lock-free parallel data structures, topological hashing for aggressive pruning, and direct rasterization optimizations, our implementation achieves more than $100\times$ speedup on large-scale benchmarks compared with the official baseline, while maintaining high accuracy.
\end{enumerate}

\section{Preliminaries and Problem Formulation}
\begin{figure}[htbp]
    \centering
    \includegraphics[width=0.95\textwidth]{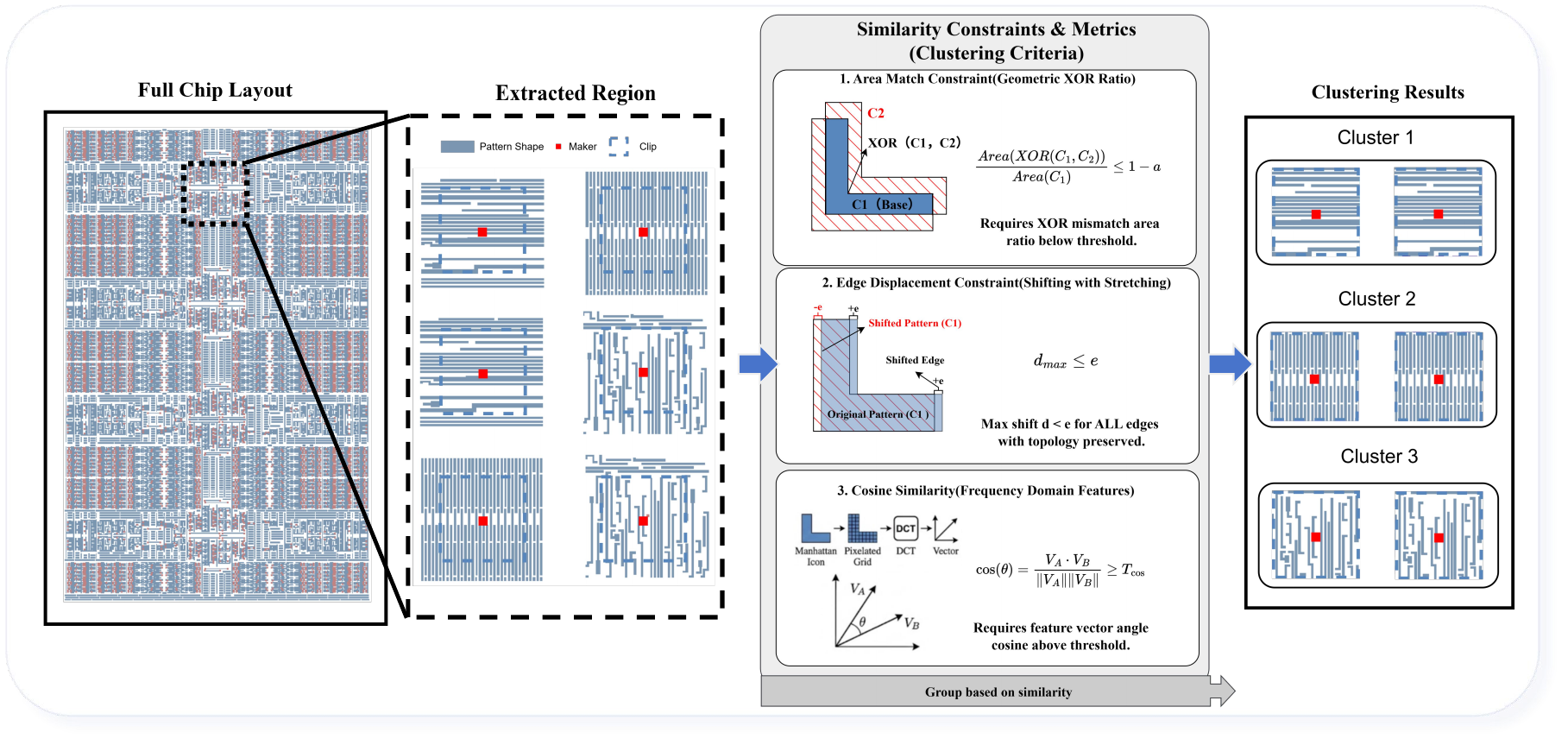} 
    \caption{Overview of the clustering workflow and similarity constraints.}
    \label{fig:constraints}
\end{figure}
\subsection{Data Representation}
The input layout data consists of two specific layers \cite{contest_spec}:
\begin{itemize}
    \item \textbf{Design Layer (Layer 1/0):} This layer contains a large number of \textit{Manhattan shapes}, where all polygon edges are strict axis-aligned. These shapes represent the physical circuit patterns.
    \item \textbf{Marker Layer (Layer 2/0):} This layer contains a set of rectangular regions, denoted as Markers. Each Marker defines the valid selection range for a pattern's center point.
\end{itemize}

A \textit{Layout Pattern} $P_i$ is defined as the geometric content on the Design Layer enclosed within a square window of a fixed \textit{Pattern Radius} ($R$), centered at a specific coordinate $c_i$. A critical property of this problem is that the center $c_i$ is not pre-determined; it must be selected from within the corresponding Marker region $m_i$:
\begin{equation}
    c_i \in m_i, \quad P_i = \text{Extract}(\text{DesignLayer}, c_i, R)
\end{equation}

\subsection{Similarity Constraints}
Two patterns $P_i$ and $P_j$ are considered clusterable if they satisfy the specific similarity constraint. We consider two distinct constraint scenarios:

\subsubsection{Cosine Similarity Constraint}
This constraint handles pixel-level similarity through frequency domain analysis. The evaluation process is defined as follows:
\begin{enumerate}
    \item \textbf{Rasterization:} Convert the vector-based patterns $P_i$ and $P_j$ into bitmap images.
    \item \textbf{DCT:} Extract frequency domain feature vectors ($A$ and $B$) using Discrete Cosine Transform.
    \item \textbf{Similarity Check:} Calculate the cosine similarity. The patterns are considered similar if the result exceeds the threshold $T_{cos}$:
    \begin{equation}
        sim(A,B) = \frac{\sum A_i \times B_i}{\sqrt{\sum A_i^2} \times \sqrt{\sum B_i^2}} > T_{cos}
    \end{equation}
\end{enumerate}

\subsubsection{Edge Movement Constraint}
This constraint enforces geometric topology consistency. Two patterns are considered similar only if they meet two conditions:
\begin{enumerate}
    \item \textbf{Polygon Overlap:} For the pattern with fewer polygons, each of its polygons must overlap with \textit{exactly one} polygon in the other pattern.
    \item \textbf{Edge Offset:} The displacement between any pair of corresponding edges must not exceed the specified threshold $T_{edge}$.
\end{enumerate}

\subsection{Problem Definition}
The objective of the \textit{Ultra-Large Scale Layout Pattern Clustering} problem is to partition $N$ Markers into a minimum number of clusters. Unlike standard clustering tasks, this problem requires determining both the \textit{cluster assignment} and the \textit{optimal center location} simultaneously.

Formally, given a set of Markers $\mathcal{M} = \{m_1, \dots, m_N\}$, we aim to find a set of centers $\mathcal{C} = \{c_1, \dots, c_N\}$ and a partition $\mathcal{K}$, such that:
\begin{enumerate}
    \item \textbf{Center Validity:} Each center $c_i$ is located within its marker $m_i$.
    \item \textbf{Cluster Consistency:} All patterns in a cluster must satisfy the designated constraint (Cosine or Edge Movement) with the cluster's representative pattern.
    \item \textbf{Minimization:} The objective is to minimize the total number of clusters $|\mathcal{K}|$.
\end{enumerate}

It is observed that defaulting the pattern center to the geometric center of the Marker typically yields sub-optimal solutions. Therefore, the core challenge lies in searching for the optimal $c_i$ within the continuous Marker region to maximize pattern similarity.

\section{Methodology: The Double-Track Framework}
\begin{figure}[htbp]
    \centering
    \includegraphics[width=0.95\textwidth]{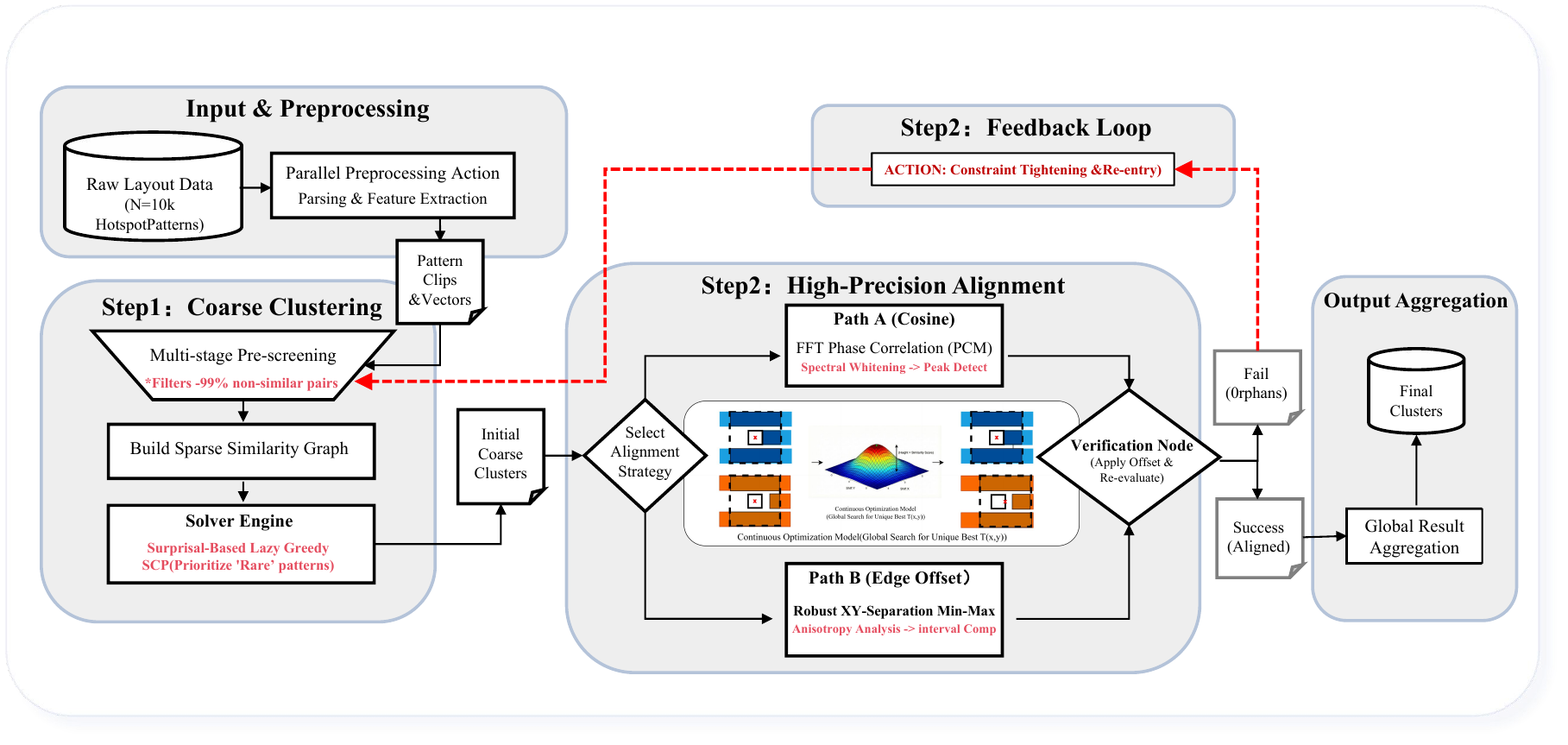} 
    \caption{Architecture of the proposed Optimal Alignment-Driven Iterative Closed-Loop Framework.}
    \label{fig:framework}
\end{figure}
\subsection{Framework Overview}
To overcome the computational prohibitiveness of $O(N^2)$ pairwise comparisons and the accuracy limitations caused by blind center selection, we propose an \textbf{Optimal Alignment-Driven Iterative Closed-Loop Framework}. 
As illustrated in Figure \ref{fig:framework}, this framework adopts a \textit{"Double-Track Parallel, Iterative Refinement"} strategy.

Depending on the input constraint type (Cosine Similarity or Edge Displacement), the system automatically branches into one of two optimized execution tracks. While the underlying alignment kernels differ (utilizing FFT for cosine and Geometric Min-Max for edge displacement), both tracks share a unified \textbf{"Coarse-to-Fine"} iterative workflow designed to achieve the optimal trade-off between execution speed and solution quality. The core workflow consists of four sequential stages that form a closed feedback loop:

\begin{enumerate}
    \item \textbf{High-Speed Pre-screening (Filtering):} 
    Handling ultra-large-scale data requires aggressive pruning of the search space. Before any expensive geometric or frequency-domain comparison, we apply a lightweight pre-screening mechanism based on topological hashing or low-resolution features. This step effectively filters out over 99\% of non-similar pairs, reducing the computational complexity from quadratic to near-linear.
    
    \item \textbf{Coarse Clustering via Surprisal-Based SCP:} 
    In this stage, to capture all potentially similar patterns, we first apply \textbf{relatively relaxed constraints} (e.g., loosened similarity thresholds) for initial screening. Based on these candidate pairs that have the potential to meet the similarity requirements, we construct a sparse similarity graph and model the clustering task as a \textbf{Set Cover Problem (SCP)}. Unlike traditional methods that rely on static greedy approaches, we employ a \textbf{Surprisal-Based Lazy Greedy} heuristic. This solver prioritizes "rare" patterns (those with high information content) to generate a set of initial, high-confidence candidate clusters in millisecond-level time \cite{adamo2023surprisal}.
    
    \item \textbf{Refinement via Optimal Alignment:} 
    The initial clusters produced in the coarse stage may contain members that are not yet perfectly aligned. In this refinement stage, we invoke our \textbf{Optimal Alignment Engine} (detailed in Section 3.2) to calculate the global optimal translation vector $T_{opt}$ of the \textbf{cluster member} relative to the \textbf{fixed cluster representative}. This operation aims to achieve optimal geometric matching, thereby \textbf{maximizing the similarity} between the member and the representative. A pattern is accepted into the cluster only if its similarity, after being significantly improved by applying $T_{opt}$, satisfies the strict constraint thresholds. 
    
    \item \textbf{Iterative Re-clustering (Feedback Loop):} 
    Patterns that fail the rigorous refinement step are not discarded but are identified as "orphans" (or outliers). These patterns are re-injected into the processing queue for the next iteration. In subsequent rounds, the system may adjust constraint tightness or search strategies to handle these "stubborn" patterns.
\end{enumerate}

This closed-loop mechanism ensures \textbf{convergence}: the iteration continues until every pattern is uniquely assigned to a cluster. By continuously stripping away easy-to-match patterns in early iterations and focusing computational resources on difficult patterns in later stages, our framework achieves a \textbf{93.4\% compression rate} while maintaining efficient runtime performance, successfully solving the NP-Hard problem with high precision.

\subsection{Optimal Alignment Engine: Analytical Geometry and Frequency Domain Optimization}
\begin{figure}[htbp]
    \centering
    \includegraphics[width=0.85\textwidth]{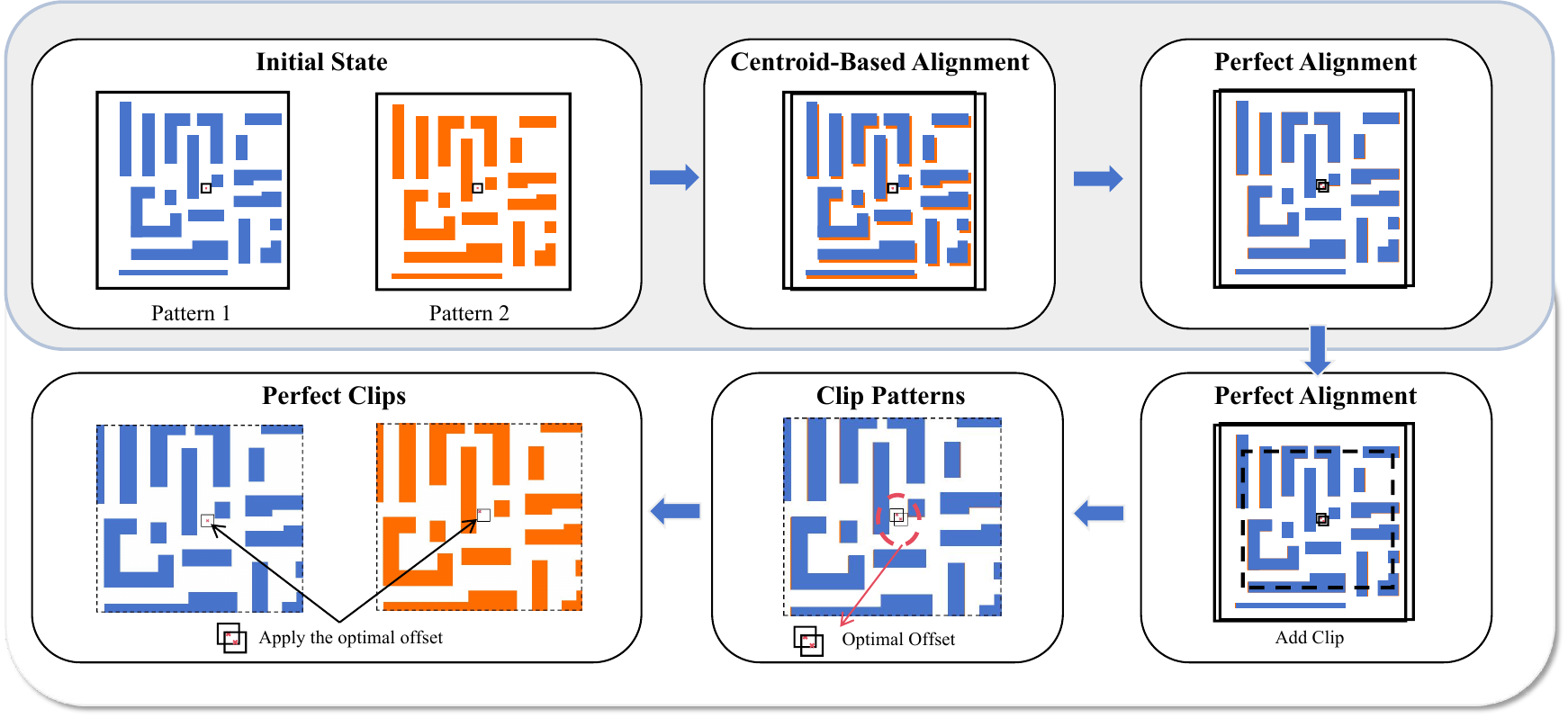} 
    \caption{Illustration of the optimal alignment process and similarity maximization. Unlike centroid-based alignment, optimal alignment actively searches for the translation vector $T_{opt}$ that maximizes the overlap between patterns.}
    \label{fig:alignment_process}
\end{figure}
The challenge of pattern clustering lies in the uncertainty of the pattern center. As demonstrated in Figure \ref{fig:alignment_process}, relying solely on geometric centroids often leads to sub-optimal similarity scores, causing geometrically similar patterns to be misclassified as distinct clusters. To overcome this, it is essential to compute an optimal translation vector $T_{opt}$ in the continuous domain that strictly maximizes the similarity(e.g., XOR overlap or Cosine score) between the candidate pattern and the cluster representative.

Our alignment engine offers analytically derived solutions for both cosine-based and edge-displacement-based constraints to solve this problem efficiently.
\subsubsection{Alignment Strategies for Area-based (Cosine) Constraints}

\begin{figure}[htbp]
    \centering
    \includegraphics[width=0.95\textwidth]{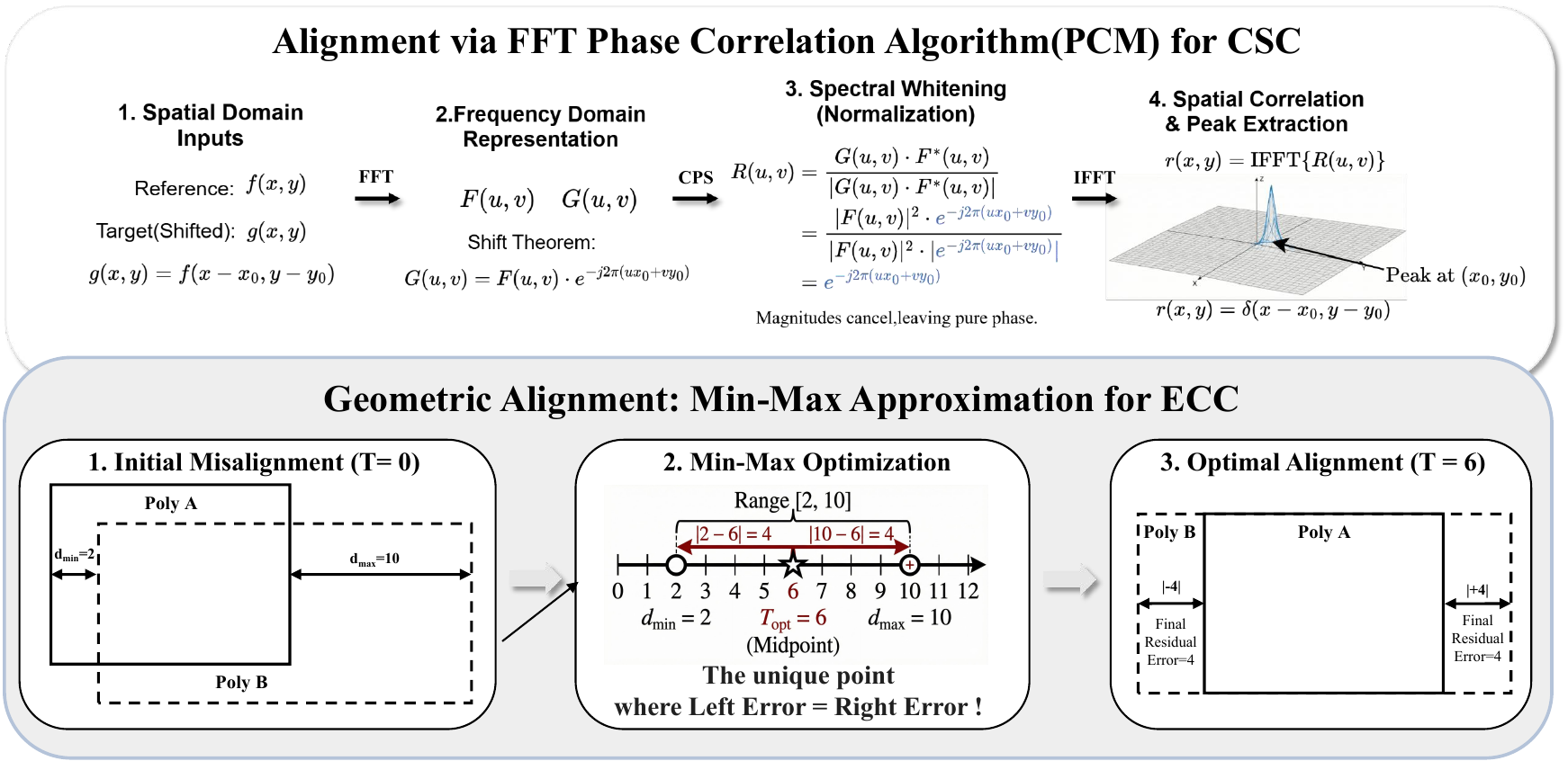} 
    \caption{Schematic of the proposed high-precision (FFT) and high-performance (Geometric) alignment workflows.}
    \label{fig:alignment_strategies}
\end{figure}
To address the alignment challenge under different performance requirements, we introduce two distinct strategies. \textbf{Figure \ref{fig:alignment_strategies} illustrates the comprehensive workflows of our approach:} the FFT-based method decouples spatial features for maximum precision (Top), while the Geometric Min-Max strategy utilizes analytical interval optimization for extreme speed (Bottom). The mathematical derivations for these strategies are detailed below.

\paragraph{(1) High-Precision: FFT-based Phase Correlation (PCM)}
To decouple image content from its spatial offset, we adopt the \textit{Fourier Shift Theorem}. Let $f(x,y)$ be a reference pattern and $g(x,y)=f(x-x_0,y-y_0)$. Their frequency-domain relationship is:
\begin{equation}
    G(u,v)=F(u,v)e^{-j2\pi(ux_0+vy_0)}.
\end{equation}

To remove image content magnitude, we compute the normalized cross-power spectrum:
\begin{equation}
    R(u,v)=\frac{G(u,v)F^*(u,v)}{|G(u,v)F^*(u,v)|}
        =e^{-j2\pi(ux_0+vy_0)}.
\end{equation}

Applying IFFT yields:
\begin{equation}
    r(x,y)=\text{IFFT}\{R(u,v)\}=\delta(x-x_0, y-y_0),
\end{equation}
where the peak location directly gives the global optimal shift $T_{opt}$, without any iterative search.

\paragraph{(2) High-Performance: Robust XY-Separation Min-Max Strategy}
For large-scale clustering, FFT can be expensive. We therefore propose a geometric alternative based on \textbf{Interval Competition}:

\begin{enumerate}
    \item Extract X/Y-axis allowable intervals $[d_{min}, d_{max}]$ for each polygon pair.
    \item Define interval quality $Q = d_{max}-d_{min}$; narrower intervals indicate stronger alignment constraints.
    \item Select axis-wise winners:
    \[
        k_x^*=\arg\min_k (d_{max,x,k}-d_{min,x,k}),\quad
        k_y^*=\arg\min_k (d_{max,y,k}-d_{min,y,k}).
    \]
    \item Analytical synthesis:
    \[
        T_{opt}=\left(
        \frac{d_{min,x,k_x^*}+d_{max,x,k_x^*}}{2},
        \frac{d_{min,y,k_y^*}+d_{max,y,k_y^*}}{2}
        \right).
    \]
\end{enumerate}

This method runs over $6\times$ faster than FFT while maintaining stable accuracy.

\subsubsection{Alignment Strategy for Edge Displacement Constraints: Geometric Min-Max Approximation}
\begin{figure}[htbp]
    \centering
    \includegraphics[width=0.85\textwidth]{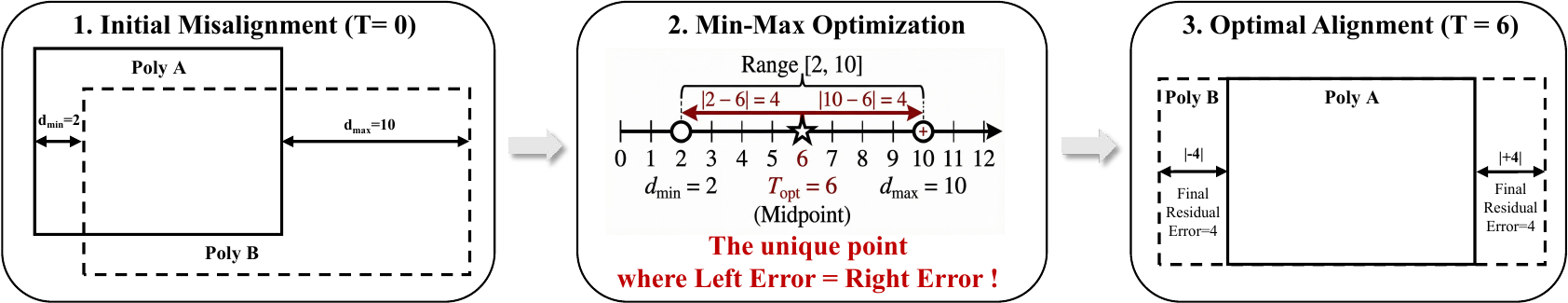} 
    \caption{Demonstration of the Geometric Min-Max alignment strategy.}
    \label{fig:ecc_alignment}
\end{figure}
For strict Edge Displacement constraints, the objective is to find a translation $T$ that minimizes the maximum deviation for all edges, mathematically formulated as a Min-Max optimization problem:
\begin{equation}
    T_{opt}=\arg\min_T \max_i \|d_i - T\|_\infty.
\end{equation}

Since the $L_\infty$ norm is separable, we can decompose the problem into independent 1D problems for X and Y axes. As illustrated in the graphical example in \textbf{Figure \ref{fig:ecc_alignment}}, the optimization process proceeds in three analytical steps:

\begin{enumerate}
    \item \textbf{Interval Extraction (Step 1):} First, the algorithm scans all edge pairs to determine the global feasible translation interval. In the example shown, the feasible range is identified as $[d_{min}, d_{max}] = [2, 10]$.
    
    \item \textbf{Midpoint Balancing (Step 2):} To minimize the maximum residual error (the worst-case scenario), the optimal strategy is to equilibrate the deviations from the interval boundaries. The unique global optimum $T_{opt}$ is derived analytically as the geometric midpoint:
    \[
        T_{opt} = \frac{d_{min} + d_{max}}{2} = \frac{2 + 10}{2} = 6.
    \]
    As visualized in the number line, this point ensures that the "Left Error" ($|2-6|=4$) strictly equals the "Right Error" ($|10-6|=4$). Any deviation from this midpoint would inevitably increase the error on one side.
    
    \item \textbf{Optimal State (Step 3):} By applying the translation $T_{opt}=6$, the patterns achieve the optimal alignment state where the residual edge displacement is minimized to $E_{min}=4$. 
\end{enumerate}

This geometric approach allows us to solve for the global optimum in $O(N)$ time (linear scan for min/max) without any iterative search, significantly outperforming traditional heuristic methods.

\subsection{Surprisal-Based SCP Solver with Lazy Acceleration}
To minimize the number of clusters required to cover all layout patterns, we formulate the clustering task as a \textbf{Set Cover Problem (SCP)}. While the classical Chvátal greedy algorithm is computationally lightweight, it suffers from a critical limitation: it treats all uncovered elements equally, ignoring the non-uniform distribution of layout data. In VLSI layouts, "rare" or isolated patterns are significantly harder to cover than frequent ones and are more prone to becoming outliers.

Drawing inspiration from the \textit{Surprisal-Based Greedy Heuristic (SBH)}, we incorporate concepts from \textbf{Information Theory} into our solver. The core intuition is to prioritize covering "High Surprisal" (rare) patterns first, as they possess higher information content and have a lower probability of being covered incidentally in later iterations.

\subsubsection{Surprisal Metric and Scoring Function}
Unlike the logarithmic self-information used in classical surprisal formulations, we design a simplified reciprocal frequency metric to meet the high-throughput requirements of ultra-large-scale data.

We define the \textbf{Graph Frequency} $f_i$ of a pattern $i$ as its degree (number of neighbors) in the sparse similarity graph:
\begin{equation}
    f_i = 1 + |neighbors(i)|
\end{equation}

The \textbf{Surprisal} $S_i$ of pattern $i$ is defined as the inverse of its frequency:
\begin{equation}
    S_i = \frac{1}{f_i}
\end{equation}

Intuitively, a smaller $f_i$ implies that pattern $i$ is rarer, resulting in a higher surprisal value $S_i$.

Based on this, we define the \textbf{Score Function} for a candidate cluster representative $j$. Instead of merely maximizing the count of covered elements, our solver maximizes the \textbf{total information content} of the covered set:
\begin{equation}
    Score_j = S_j + \sum_{k \in neighbors(j) \cap U} S_k
\end{equation}
where $U$ denotes the set of currently uncovered patterns. This formulation directs the solver to select representatives that cover "stubborn" or "rare" nodes, effectively reducing the total cluster count.

\subsubsection{Lazy Greedy Acceleration}
A standard greedy approach requires updating the scores of all remaining candidates after every selection, incurring a prohibitive cost of $O(N \cdot d_{avg})$ per step. To handle millions of patterns efficiently, we integrate a \textbf{Lazy Greedy} mechanism that exploits the \textit{submodularity} of the scoring function:

\begin{enumerate}
    \item \textbf{Initialization:} Compute initial scores for all candidates and maintain them in a Max-Priority Queue.

    \item \textbf{Lazy Evaluation Strategy:}
    \begin{itemize}
        \item Pop the top candidate $j_{top}$ with the highest tentative score.
        \item \textbf{Staleness Check:} If the score is outdated (i.e., some neighbors of $j_{top}$ have been covered by previously selected centers), we strictly re-calculate its \textbf{Real Marginal Gain} based on the current uncovered set $U$.
        \item \textbf{Select or Re-insert:}  
        If the re-calculated score remains higher than the new top of the heap, $j_{top}$ is guaranteed to be the optimal choice (due to submodularity) and is selected immediately.  
        Otherwise, $j_{top}$ is re-inserted into the queue with its updated score.
    \end{itemize}
\end{enumerate}

This strategy avoids the vast majority of redundant score updates. Experimental results demonstrate that this optimization reduces the amortized complexity of a single iteration to approximately
\[
O(d_{avg} \cdot \log N),
\]
achieving over \textbf{20$\times$ speedup} compared to the standard implementation and enabling millisecond-level convergence.

\section{Experimental Results}

\subsection{Experimental Setup}
All experiments were conducted on the official contest server powered by a \textbf{HiSilicon Kunpeng-920} processor (ARMv8 architecture, 3.0 GHz) equipped with 128 physical cores and 64GB of memory. To ensure fair evaluation and benchmark consistency, all algorithms were strictly pinned to run with \textbf{32 threads} (utilizing 32 physical cores).

\subsection{Overall Performance: Comparison with Official Baseline}
To validate the effectiveness of our framework, we compared our solution against the official contest baseline across various scales and constraint scenarios. Table \ref{tab:baseline_comparison} presents the detailed results.

\begin{table}[h]
\centering
\caption{Comparison of Cluster Count and Runtime between Baseline and Ours}
\label{tab:baseline_comparison}

\resizebox{\textwidth}{!}{%
\begin{tabular}{lcccccc}
\toprule
\multirow{2}{*}{\textbf{Test Case (Scale)}} & \multicolumn{3}{c}{\textbf{Cluster Count (Lower is Better)}} & \multicolumn{3}{c}{\textbf{Runtime (Lower is Better)}} \\ 
\cmidrule(lr){2-4} \cmidrule(lr){5-7}
 & \textbf{Baseline} & \textbf{Ours} & \textbf{Imp. ($\times$)} & \textbf{Baseline} & \textbf{Ours} & \textbf{Imp. ($\times$)} \\ \midrule

Case 1 ($N=20$) & 13 & \textbf{1} & 13.0$\times$ & 500ms & \textbf{100ms} & 5.0$\times$ \\
Case 2 ($N=20$) & 16 & \textbf{2} & 8.0$\times$ & 50ms & \textbf{7ms} & 7.1$\times$ \\ \midrule
Case 3 ($N=10k$) & 3500 & \textbf{660} & 5.3$\times$ & 12500s & \textbf{99s} & \textbf{126.3$\times$} \\
Case 4 ($N=10k$) & 6500 & \textbf{2730} & 2.4$\times$ & 12500s & \textbf{7s} & \textbf{178.6$\times$} \\
Case 5 ($N=10k$) & 3500 & \textbf{658} & 5.3$\times$ & 12500s & \textbf{99s} & \textbf{126.3$\times$} \\
Case 6 ($N=10k$) & 6500 & \textbf{2626} & 2.5$\times$ & 12500s & \textbf{7s} & \textbf{178.6$\times$} \\ 
\bottomrule

\multicolumn{7}{l}{\footnotesize \textbf{Note:} "Imp." denotes the improvement ratio calculated as $\mathbf{Baseline / Ours}$.} \\ 
\end{tabular}%
}
\end{table}

Key conclusions drawn from Table \ref{tab:baseline_comparison} include:
\begin{enumerate}
    \item \textbf{Extreme Speedup ($>100\times$):} In the challenging large-scale cases (Case 3--6), the baseline requires over 3.5 hours, whereas our solution finishes in seconds (7s--99s). Notably, in Edge Displacement scenarios (Case 4/6), we achieve a massive \textbf{178.6$\times$} speedup.
    \item \textbf{Superior Compression (93.4\% Reduction):} Our clustering quality significantly outperforms the baseline. For instance, in Case 3, the cluster count drops from 3500 to 660, representing a 5.3$\times$ improvement. Crucially, relative to the raw input of 10,000 unique patterns, our approach achieves a maximum \textbf{93.4\% Cluster Reduction}, demonstrating the capability of our Optimal Alignment Engine to identify potential similarities that the baseline misses.
    \item \textbf{Ranking:} Due to these advantages in both speed and quality, our solution secured the \textbf{First Place} award in \textit{Problem 7: Ultra-Large Scale Layout Pattern Clustering Algorithm}, sponsored by \textbf{HiSilicon}, at the \textit{2025 China Postgraduate EDA Elite Challenge}.
\end{enumerate}

\subsection{Ablation Study: Component-wise Optimization Analysis}
To identify the sources of our performance gains, we analyzed the acceleration contribution of each core algorithmic module. Table \ref{tab:component_analysis} summarizes the results.

\begin{table}[h]
\centering
\caption{Breakdown of Algorithmic Optimizations and Speedup Contribution}
\label{tab:component_analysis}
\resizebox{\textwidth}{!}{%
\begin{tabular}{lcccc}
\toprule
\textbf{Algorithmic Module} & \textbf{Baseline / Unoptimized} & \textbf{Optimized Method} & \textbf{Performance} & \textbf{Speedup} \\ \midrule
Cosine Similarity (CSC) & \begin{tabular}[c]{@{}c@{}}Serial Calculation\\ (Full Matrix)\end{tabular} & \begin{tabular}[c]{@{}c@{}}\textbf{Parallel + Vectorized}\\ (Rasterize-Direct)\end{tabular} & 6 min $\to$ \textbf{6s} & \textbf{60$\times$} \\ \midrule
Graph Construction & \begin{tabular}[c]{@{}c@{}}Brute-force $O(N^2)$\\ (All Pairs)\end{tabular} & \begin{tabular}[c]{@{}c@{}}\textbf{Multi-stage Pre-screening}\\ (\textbf{99\% Filtered})\end{tabular} & Hours $\to$ \textbf{Seconds} & \textbf{>1000$\times$} \\ \midrule
SCP Solver & \begin{tabular}[c]{@{}c@{}}Standard Greedy\\ (Full Update)\end{tabular} & \begin{tabular}[c]{@{}c@{}}\textbf{Surprisal Lazy Greedy}\\ (Lazy Evaluation)\end{tabular} & 1s $\to$ \textbf{50ms} & \textbf{20$\times$} \\ \midrule
Overall Computing & Single Thread & \textbf{32 Threads Parallel} & - & \textbf{12$\times$} \\ \bottomrule
\end{tabular}%
}
\end{table}

As shown in Table \ref{tab:component_analysis}:
\begin{itemize}
    \item \textbf{Bottleneck Breaker ($>1000\times$):} The \textit{Multi-stage Pre-screening} effectively filters over 99\% of redundant pairs, reducing Graph Construction time from hours to seconds.
    \item \textbf{Core Acceleration (60$\times$):} \textit{Direct Rasterization} with SIMD vectorization significantly boosts the throughput of the Cosine Similarity check.
    \item \textbf{Solver Efficiency (20$\times$):} The \textit{Lazy Greedy} strategy reduces the SCP solving overhead by avoiding redundant score updates.
\end{itemize}

\section{Conclusion}
This paper presents an \textbf{Optimal Alignment-Driven Iterative Closed-Loop Framework} designed to systematically address the critical scalability and precision challenges in ultra-large-scale VLSI layout pattern clustering. The framework establishes a hierarchical \textbf{"Coarse-to-Fine"} processing paradigm: it initializes by efficiently generating \textbf{initial clustering results with alignment potential under relaxed constraints}, utilizing multi-stage pre-screening and an information-theoretic \textbf{Surprisal-Based Lazy Greedy SCP solver}. Subsequently, leveraging our \textbf{optimal alignment algorithms} (FFT-based Phase Correlation and Geometric Min-Max strategies), it achieves high-performance optimal alignment between cluster members and representatives, thereby \textbf{maximizing the similarity potential} of each pair. The iterative refinement and convergence mechanism applied to the remaining patterns achieves \textbf{extreme clustering compression} while maintaining \textbf{linear time complexity} suitable for scalability to ultra-large-scale layouts. Empirical evaluations on the 2025 China Postgraduate EDA Elite Challenge benchmarks substantiate the framework's efficacy, demonstrating a remarkable \textbf{93.4\% cluster compression ratio} relative to raw inputs and an acceleration exceeding \textbf{100$\times$} over the official baseline. These results, which secured the \textbf{First Place} award in the HiSilicon-sponsored \textit{Problem 7}, validate the solution's superior robustness and potential for advancing industrial Design for Manufacturability (DFM) applications.
\bibliographystyle{unsrt}
\bibliography{refs}

\begin{thebibliography}{1}

\bibitem{chang2017iclaire}
Wen-Hao Chang, Iris Hui-Ru Jiang, Yu-Ting Yu, and Wen-Feng Liu.
\newblock iclaire: A fast and general layout pattern classification algorithm.
\newblock In {\em Design Automation Conference (DAC)}, pages 1--6. ACM/IEEE, 2017.

\bibitem{chen2017minimizing}
Kuan-Jung Chen, Yi-Kuan Chuang, Bo-Yuan Yu, and Shao-Yun Fang.
\newblock Minimizing cluster number with clip shifting in hotspot pattern classification.
\newblock In {\em Design Automation Conference (DAC)}, pages 1--6. ACM/IEEE, 2017.

\bibitem{he2023general}
Xiao He, Yibo Wang, Zhuo-Yuan Fu, Yu-Pei Wang, and Yang Guo.
\newblock A general layout pattern clustering using geometric matching-based clip relocation and lower-bound aided optimization.
\newblock {\em ACM Transactions on Design Automation of Electronic Systems (TODAES)}, 28(6):1--23, 2023.

\bibitem{lin2023layout}
Chun-Sheng Lin, Pei-Yu Tsai, Yu-Hsun Liu, et~al.
\newblock Layout hotspot pattern clustering using a density-based approach.
\newblock {\em International VLSI Symposium on Technology, Systems and Applications (VLSI-TSA)}, pages 1--4, 2023.

\bibitem{adamo2023surprisal}
Tim Adamo, Gianpaolo Ghiani, Emanuela Guerriero, and Donato Pareo.
\newblock A surprisal-based greedy heuristic for the set covering problem.
\newblock {\em Algorithms}, 16(7):321, 2023.

\bibitem{contest_spec}
Problem 7: Ultra-large scale layout pattern clustering algorithm.
\newblock Contest Specification, 2025.
\newblock Accessed: 2025-12-09.

\end{thebibliography}

\end{document}